\documentclass[journal,12pt,onecolumn,draftclsnofoot,]{IEEEtran}
\usepackage[ruled, vlined]{algorithm2e}
\usepackage{algpseudocode}
\usepackage{amsmath}
\usepackage{amssymb}
\usepackage{array}
\usepackage{arydshln}
\usepackage{blkarray, bigstrut}
\usepackage{bm}
\usepackage{booktabs}
\usepackage{braket}
\usepackage{cite}
\usepackage{cuted}
\usepackage{enumitem}
\usepackage{graphicx}
\usepackage[mathscr]{euscript}
\usepackage{mathtools}
\usepackage{multirow}
\usepackage{multicol}
\usepackage{framed} 
\usepackage[framed]{ntheorem}
\usepackage{nicefrac}
\usepackage{pgfplots}
\usepackage{qtree}
\usepackage{adjustbox}
\usepackage{rotating}
\pgfplotsset{compat = 1.13}
\usepackage{stfloats}
\usepackage{subfig}
\usepackage{url}
\usepackage{tikz}
\usepackage{tkz-berge}
\usepackage[framemethod=TikZ]{mdframed}
\usetikzlibrary{patterns}
\usetikzlibrary{spy}
\usetikzlibrary{shapes, backgrounds,calc, snakes}
\usetikzlibrary{decorations.pathreplacing}
\usetikzlibrary{matrix}
\usepackage{fancybox}

\tikzstyle{vertex} = [circle, draw, inner sep = 0pt, minimum size = 10pt]

\newframedtheorem{frm-prop}{Property}

\definecolor{bblue}{rgb}{0.12392, 0.0490, 0.9588}
\definecolor{sskyblue}{rgb}{0.1529, 0.5882, 0.9216}
\definecolor{ggreen}{rgb}{0.5020, 0.7961, 0.3451}
\definecolor{yyellow}{rgb}{0.9765, 0.9804, 0.0784}

\definecolor{color0}{HTML}{FF0147}
\definecolor{color1}{HTML}{F400DC}
\definecolor{color2}{HTML}{BA0DFF}
\definecolor{color3}{HTML}{5700E8}
\definecolor{color4}{HTML}{0B03FF}
\definecolor{color5}{HTML}{0957F4}
\definecolor{color6}{HTML}{03B3FF}
\definecolor{color7}{HTML}{08E8DA}
\definecolor{color8}{HTML}{07FF8E}
\definecolor{color9}{HTML}{51FF0A}

\definecolor{p1}{rgb}{1, 0.0667, 0}
\definecolor{p2}{rgb}{1, 0.24, 0}
\definecolor{p3}{rgb}{1, 0.349, 0}
\definecolor{p4}{rgb}{1, 0.490, 0}
\definecolor{p5}{rgb}{1, 0.631, 0}
\definecolor{p6}{rgb}{1, 0.792, 0}
\definecolor{p7}{rgb}{1, 0.933, 0}

\begin{document}

\title{Subchannel Allocation for Vehicle--to--Vehicle Broadcast Communications in Mode-3}

\author{Luis F.~Abanto-Leon, Arie Koppelaar, Sonia Heemstra de Groot}

\maketitle

\begin{abstract}
	Conversely to mainstream cellular networks where uplink / downlink data traffic is centrally managed by eNodeBs, in vehicle--to--vehicle (V2V) broadcast communications \textit{mode-3} eNodeBs engage solely in subchannel assignment but ultimately do not intervene in data traffic control. Accordingly, vehicles communicate directly with their counterparts utilizing the allotted subchannels. Due to its loosely controlled one--to--all nature, V2V \textit{mode-3} is advantageous for time-critical applications. Nevertheless, it is imperative that the assignment of subchannels is accomplished without conflicts while at the same time satisfying quality of service (QoS) requirements. To the best of our knowledge, there exists no unified framework for V2V \textit{mode-3} that contemplates both prevention of allocation conflicts and fulfillment of QoS. Thus, four types of conditions that are of forceful character for attaining QoS-aware conflict-free allocations have been identified: $(i)$ assure differentiated QoS per vehicle, $(ii)$ preclude intra-cluster subframe conflicts, $(iii)$ secure minimal time dispersion of allotted subchannels and $(iv)$ forestall one-hop inter-cluster subchannel conflicts. Such conditions have been systematized and merged in an holistic manner allowing non-complex manipulation to perform subchannel allocation optimization. In addition, we propose a surrogate relaxation of the problem that does not affect optimality provided that certain requisites are satisfied.
\end{abstract}

\begin{IEEEkeywords}
	subchannel allocation, broadcast vehicular communications, mode-3, sidelink
\end{IEEEkeywords}

\IEEEpeerreviewmaketitle

\section{Introduction}
The 3rd Generation Partnership Project (3GPP) \cite{b2} recently outlined two kinds of resource allocation notions for vehicle--to--vehicle (V2V) broadcast communications. One of these schemes is known as V2V \textit{mode-3} and requires the involvement of eNodeBs to realize subchannel\footnote{A subchannel is a time-frequency chunk consisting of a number of resource blocks (RBs) \cite{b1}.} assignment to vehicles in coverage\footnote{The other scheme, known as V2V \textit{mode-4}, is not focus of this paper but subchannel allocation is contrastingly attained on a distributed basis, i.e. without the intervention of eNodeBs. Thus, a vehicle primarily senses the energy levels across subchannels to discern between occupancy and utilization. Thereupon, from a list of potentially unoccupied subchannels, each vehicle self-allocates a suitable one for its own usage while making an effort not to generate conflicts or compromise the link reliability of other vehicles \cite{b6}.}. In \textit{mode-3}, eNodeBs participate limitedly only assigning subchannels to vehicles in coverage. The assignment of subchannels is accomplished taking into consideration that $(i)$ vehicles are dispersed over several communications clusters and that $(ii)$ some subchannels can be reused provided that the interference generated due to repurposing is controlled \cite{b2} \cite{b5}. Subsequently, the allocation of subchannels is notified via downlink to all vehicles in coverage, which will engage in direct communications with their counterparts. Typically, the kind of message that vehicles would exchange are fixed-size cooperative awareness messages (CAMs), which convey information about the speed, position, direction, etc. of each vehicle. However, depending on the application to be supported, a greater number of subchannels might be required by each vehicle \cite{b3}. On the other hand, certain assignments of subchannels may result in vehicles conflicting with each other, e.g. when two vehicles in the same cluster transmit concurrently in the same subframe albeit different subchannels\footnote{Although vehicles might be transmitting in distinct subchannels of the same subframe, this case is deemed as an incongruity due to half-duplex PHY assumption which allows vehicles to either transmit or receive at a time.}. In an effort to attain a conflict-free allocation and provide each vehicle with the required quality of service (QoS), we have identified four types of conditions that should be satisfied. In addition, a subchannel allocation framework that includes the aforementioned conditions is developed.

Our paper is structured as follows. Section II justifies the motivation of this work and describes our contributions. Section III elaborates on a proposed framework for the subchannel allocation problem in V2V \textit{mode-3} considering four types of conditions. Section IV elaborates on a relaxed version of the problem presented in Section III. Section V discusses simulation results whereas Section VI summarizes the conclusions of this work.

\section{Motivation and Contributions}
\begin{figure*}
	\begin{center}
		\begin{tikzpicture}
			\node (img) {\includegraphics[width=0.85\linewidth]{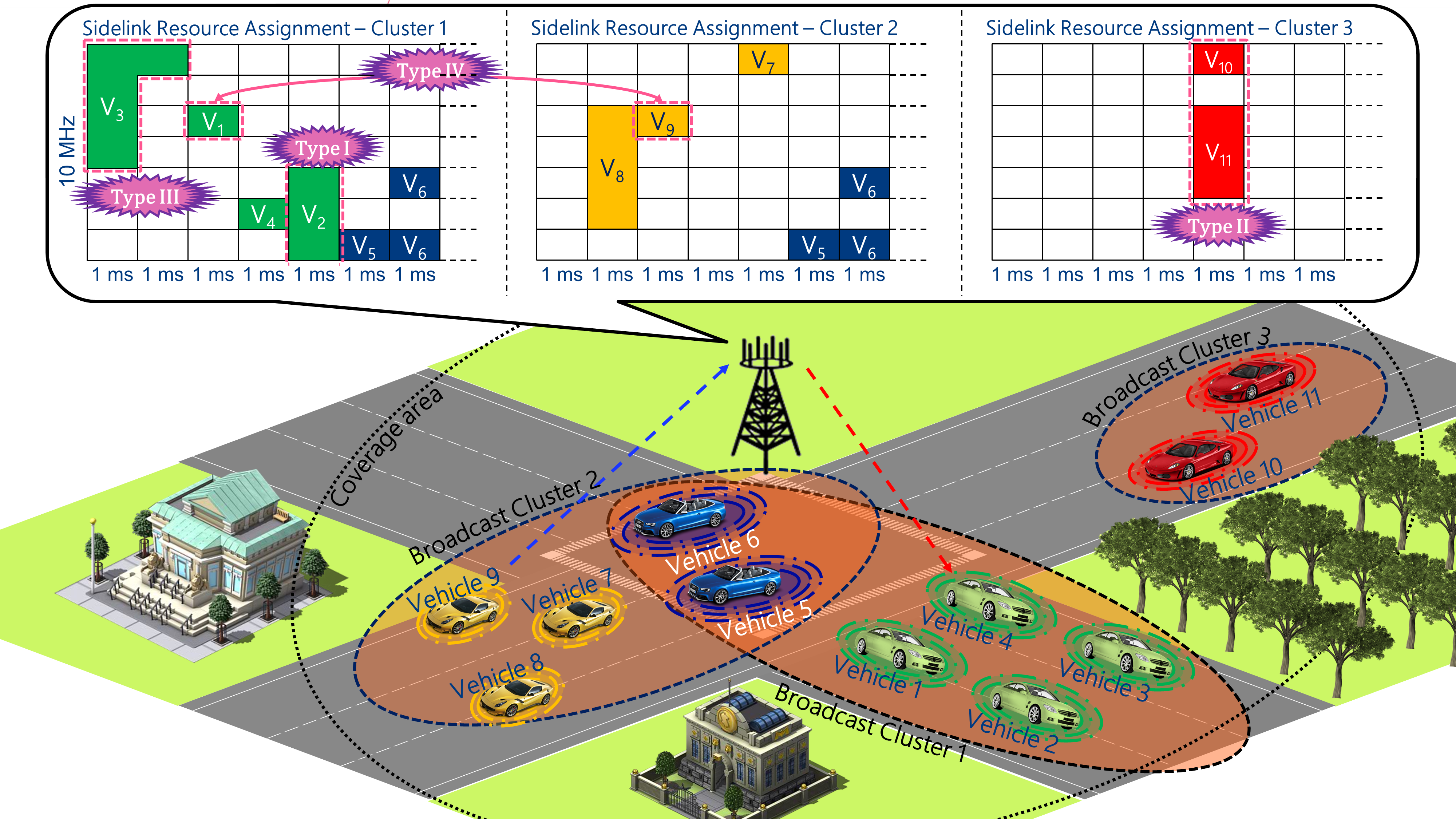}};
		\end{tikzpicture}
		\caption{Scenario of vehicular broadcast communications in \textit{mode}-3}
		\label{f1}
		\vspace{-0.5cm}
	\end{center}
\end{figure*}

A typical vehicular scenario is shown in Fig. \ref{f1}, wherein $N = 11$ vehicles are distributed over 3 clusters. To wit, \textit{cluster 1} consists of $N_1 = 6$ vehicles, namely $\{ v_1, v_2, v_3, v_4, v_5, v_6 \}$; \textit{cluster 2} consists of $N_2 = 5$ vehicles, namely $\{ v_5, v_6, v_7, v_8, v_9 \}$ whereas \textit{cluster 3} is constituted by $N_3 = 2$ vehicles, i.e. $\{ v_{10}, v_{11} \}$. Furthermore, at the intersection of \textit{cluster 1} and \textit{cluster 2} lie vehicles $\{ v_5, v_6 \}$. Depending on the distribution of vehicles and their affiliation with the different clusters, some subchannel assignments might be detrimental as one or more types of undesired effects might be generated thus impinging on reliability. Accordingly, we identify four types of conditions that are necessary for QoS-aware conflict-free allocations.

\begin{itemize}
	\item \textit{Type I}: There is a per-vehicle QoS requirement. In this paper, we define QoS in terms of the channel capacity required by a vehicle to broadcast its signal. For instance, in Fig. \ref{f1}, vehicle $v_2$ requires 3 subchannels. 
	\item \textit{Type II}: When two vehicles in the same cluster broadcast their respective signals simultaneously, they will not be able to receive the signal of the other. This problem will only affect vehicles transmitting concurrently. However, other vehicles in the cluster will be able to receive the signals and decode them if the received power is sufficiently high and the signals were transmitted in non-overlapping subchannels. This situation is depicted by $v_{10}$ and $v_{11}$, which broadcast in subchannels of the same subframe.
	\item \textit{Type III}: It is important that each vehicle transmits the intended signal using solely subchannels contained within the same subframe (1 ms duration). This will thereby prevent resource spreading over time, which may be detrimental---specially in highly congested scenarios---since the maximum number of vehicles that could be supported in time would decrease. This problem is depicted by vehicle $v_3$ whose subchannels span two subframes.
	\item \textit{Type IV}: Specifically for vehicles that lie at the intersection of clusters, this effect is critical as they will receive concurrent signals from other vehicles that are not aware of each other. This effect is similar to the hidden node problem experienced in CSMA \cite{b4}. Conversely to Type II, in this case, signals are concurrent in time and frequency and thus they may not be decodable by other vehicles. For instance, if vehicles $v_1$ and $v_9$ transmit in the same subchannel, the affected vehicles would be $v_5$ and $v_6$ as they will receive combined signals from vehicles in distinct clusters.
\end{itemize}

Our contribution is the identification of necessary allocation conditions to prevent four types of conflicts, which are likely to occur in sidelink V2V \textit{mode-3} given the new channelization structure proposed by 3GPP. In addition, we also propose a mathematical framework to perform subchannel allocation taking into consideration the aforementioned conditions. Subsequently, we derive an exact formulation of the problem and also propose a relaxed version that can attain optimality.

\section{Subchannel Allocation Framework}
We have considered a 10 MHz channel for exclusively supporting sidelink broadcast communications among vehicles. We assume that downlink and uplink spectrum resources are available to serve for elementary periodical signaling between vehicles and eNodeBs. Thus, vehicles can report to eNodeBs via uplink the channel conditions they perceive in each of the subchannels. Based on this information, the eNodeB allots subchannels to each of the vehicles considering $(i)$ their QoS requirements and $(ii)$ their distribution across the clusters---in order not to provoke conflicts. Finally, the eNodeB notifies the vehicles of the resultant allocation via downlink.

\begin{figure}[!t]
	\centering
	\begin{tikzpicture}[scale = 1]
	\draw[step=1cm, thick] (0,0) grid (2.2,-2.2);
	\draw[step=1cm, thick] (2.8,0) grid (4,-2.2);
	\draw[step=1cm, thick] (0,-2.8) grid (2.2,-4.001);
	\draw[step=1cm, thick] (2.8,-2.8) grid (4,-4.001);
	
	\draw[fill=black] (1,-1) rectangle (2,-1.3);
	\node at (1.5,-1.15) {\textcolor{white}{\small Control}};
	\draw[pattern=crosshatch, pattern color=black] (1,-1.3) rectangle (2,-2);
	\node[fill = white] at (1.5,-1.65) {\textcolor{black}{\small Data}};
	
	\node at (2.5,0) {\dots};
	\node at (2.5,-1) {\dots};
	\node at (2.5,-2) {\dots};
	\node at (2.5,-3) {\dots};
	\node at (2.5,-4) {\dots};
	
	\node at (0, -2.4) {\vdots};
	\node at (1, -2.4) {\vdots};
	\node at (2, -2.4) {\vdots};
	\node at (3, -2.4) {\vdots};
	\node at (4, -2.4) {\vdots};
	
	\node at (2.5, -2.45) {$\ddots$};
	
	\draw[decoration={brace, raise=5pt},decorate] (1, -4.0) -- node[right=6pt] {} (0.0, -4.0);
	\draw[decoration={brace, raise=5pt},decorate] (2.0, -4.0) -- node[right=6pt] {} (1, -4.0);
	\draw[decoration={brace, raise=5pt},decorate] (4.0, -4.0) -- node[right=6pt] {} (3, -4.0);
	
	\draw[decoration={brace, raise=5pt},decorate] (0, -1) -- node[right=6pt] {} (0.0, 0);
	\draw[decoration={brace, raise=5pt},decorate] (0, -2) -- node[right=6pt] {} (0.0, -1);
	\draw[decoration={brace, raise=5pt},decorate] (0, -4) -- node[right=6pt] {} (0.0, -3);

	\draw[decoration={brace, raise=5pt},decorate] (4, -4.65) -- node[right=6pt] {} (0.0, -4.65);
	\node at (2,-5.2) {$L$ (ms)};
	
	\draw[decoration={brace, raise=5pt},decorate] (-0.7, -4) -- node[right=6pt] {} (-0.7, 0);
	\node[rotate = 90] at (-1.3,-2) {Frequency (MHz)};
	
	\node at (0.5,-4.5) {1 ms};
	\node at (1.5,-4.5) {1 ms};
	\node at (3.5,-4.5) {1 ms};
	
	\node at (0.5,-0.5) {$r_{1}$};
	\node at (0.5,-1.5) {$r_{2}$};
	\node at (0.5,-3.5) {$r_{K}$};
	
	\node at (1.5,-0.5) {$r_{K+1}$};
	\node at (1.5,-3.5) {$r_{2K}$};
	
	\node at (3.5,-3.5) {$r_{KL}$};
	
	\node[rotate = 90] at (-0.6,-0.5) {$B$};
	\node[rotate = 90] at (-0.6,-1.5) {$B$};
	\node[rotate = 90] at (-0.6,-3.5) {$B$};
	
	\end{tikzpicture}	
	\caption{Channelization for V2V communications}
	\label{f2}
	\vspace{-0.5cm}
\end{figure}
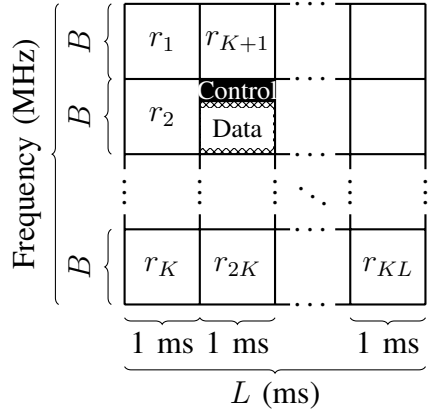

Let $L$ denote the number of subframes, each having a duration of 1 ms. In each subframe there are $K$ subchannels of bandwidth $B$ MHz, such that $KB \leq 10$ MHz, as shown in Fig. \ref{f2}. Thus, in each subframe, $K \leq 7$ subchannels can be supported. Nevertheless, depending on the context and target application, a number vehicles might require to be alloted more subchannels than others. Now, assume that $r_k$ represents a sidelink subchannel and $\mathcal{R}_l = \{ r_{(l-1)K + 1}, \dots,  r_{lK}\}$ is the set of subchannels in subframe $l$, for $l = 1, 2, \dots, L$. Hence, $\mathcal{R} = \cup_{l=1}^L \mathcal{R}_l = \{r_1, r_2, \dots, r_{KL} \} $ represents the whole set of subchannels for an allocation window of $L$ ms. Also, let $N$ denote the number of vehicles in the system, which are distributed over several clusters\footnote{Clusterization of vehicles based on similarity features such as velocity, position and direction can be advantageous in terms of simplifying the assignment of subchannels and enhancing the resources reuse ratio. In this work, such grouping is based solely on the position of vehicles.}. If $\mathcal{V}^{(j)}$ denotes a particular cluster $j$, then $\mathcal{V} = \cup_{j} \mathcal{V}^{(j)}= \{v_1, v_2, \dots, v_N \}$ is constituted by all the $N$ vehicles in the system. Furthermore, let $x_{ik} \in \rm{GF}(2)$ be a 0/1 variable that denotes the matching between vehicle $v_i \in \mathcal{V}$ and subchannel $r_k \in \mathcal{R}$. Accordingly, if $x_{ik} = 1$, then subchannel $r_k$ has been allotted to vehicle $v_i$. Conversely, if $x_{ik} = 0$, vehicle $v_i$ will not utilize $r_k$ for transmission. In addition, $c_{ik} = B \log_2(1 + \mathsf{SINR}_{ik})$ denotes the achievable capacity that vehicle $v_i$ can attain if it transmits in subchannel $r_k$. Similarly, $\mathsf{SINR}_{ik}$ is the signal--to--interference--plus--noise ratio (SINR) that vehicle $v_i$ perceives in subchannel $r_k$ \cite{b5}.

\vspace{0.1cm}
In the following, we introduce the objective function related to the subchannel allocation problem. Thereafter, the four types of assignment requirements already mentioned in Section II, namely conditions Type I, Type II, Type III and Type IV are modeled mathematically.

\vspace{0.1cm}
\underline{\textit{Objective function: }}
The objective is to maximize the system sum-capacity subject to satisfying the allocation conditions already discussed. The function to optimize can be expressed as a weighted sum ${\bf c}^T {\bf x}$, where ${\bf x}$ and ${\bf c}$ are vectors containing elements $x_{ik}$ and $c_{ik}$ for all the possible combinations of vehicles $v_i$ and subchannels $r_k$, i.e. $\mathbf{x} = [x_{1,1}, \dots, x_{1,KL}, \dots, x_{N,1}, \dots, x_{N,KL}]^T$, $\mathbf{c} = [c_{1,1}, \dots, c_{1,KL}, \dots c_{N,1}, \dots, c_{N,KL}]^T$. 

\subsection{Type I: Differentiated QoS requirements per vehicle}
We have considered per-vehicle QoS requirements in order to take into account the cases when vehicles may demand additional subchannels for conveying data---this may occur when non-safety related applications or services are targeted. Thus, for each vehicle $v_i$ the required capacity is denoted by $q_i$ and the relation satisfying such a demand is given by $ \sum_{k = 1}^{KL} c_{ik} x_{ik} = q_i$, for $i = 1, 2, \dots, N$. Since equality may not always be attained, the condition can be recast as satisfying both a lower and upper bound. As a result, we can express it as $ q_i - \epsilon \leq \sum_{k = 1}^{KL} c_{ik} x_{ik} \leq q_i + \epsilon$. For $N$ vehicles, there will be $N$ QoS-related demands to satisfy. We can express this set of conditions as follows
\begin{equation} \label{e1}
	{\bf q}_{N \times 1} - {\bm \epsilon} \leq ({\bf I}_{N \times N} \otimes {\bf 1}_{1 \times KL})({\bf c} \circ {\bf x}) \leq {\bf q}_{N \times 1} + {\bm \epsilon}
\end{equation}
where ${\bf q} = [q_1, q_2, \dots, q_N]^T$ and ${\bm \epsilon} = \epsilon \cdot ({\bf 1}_{N \times 1}) $, $\exists ~\epsilon \geq 0$. The symbols $\otimes$ and $\circ$ represent the Kronecker and Hadamard product, respectively. 

\subsection{Type II: Intra-cluster subframe allocation conflicts}
This conflict manifests when two or more vehicles in the same cluster (therefore intra-cluster) transmit in subchannels that belong to the same subframe. Under this condition, vehicles are not capable of receiving each other's signals since half-duplex interfaces are assumed by default. However, other vehicles will not have difficulty in receiving signals originated at the conflicting vehicles. In synthesis, if we can guarantee that no pair of intra-cluster vehicles will transmit concurrently in the same subframe, the conflicts can be prevented. Let $v_y$ and $v_z$ be vehicles that belong to the same cluster. Thus, to prevent multiple transmissions in the same subframe, the condition $\left( \sum_{k_1} x_{y{k_1}} \right) \left( \sum_{k_2} x_{z{k_2}} \right) = 0$ must be guaranteed. Note that the indexes  $k_1$ and $k_2$ are used to represent the subchannels $r_{k_1} \in \mathcal{R}_l$ and $r_{k_2} \in \mathcal{R}_l$, for $l = 1, 2, \dots, L$. On closer inspection, the equality above is non-zero only if vehicle $v_y$ is assigned at least one subchannel $k_1$ from subframe $\mathcal{R}_l$ and vehicle $v_z$ is assigned at least one subchannel $k_2$ from the same subframe $l$. Assuming a total number of $N$ vehicles, a compact form of expressing these constraints is given by  
\begin{equation} \label{e2}
	[({\bf G}_{P \times N}^{+} \otimes {\bf I}_{L \times L}) {\bf x}_s] \circ [({\bf G}_{P \times N}^{-} \otimes {\bf I}_{L \times L}) {\bf x}_s] = {\bf 0}_{PL \times 1}
\end{equation}
where ${\bf x}_s = ({\bf I}_{NL \times NL} \otimes {\bf 1}_{1 \times K}) {\bf x}$. The parameter $P$ is the total number of vehicle pairs across all the clusters---a pair can only be formed with vehicles from the same cluster. The matrices ${\bf G}^{+}$ and ${\bf G}^{-}$ contain information about the forbidden allocations based on the topology of the scenario and distribution of vehicles in the clusters. In Section IV, we illustrate the physical meaning of these matrices with an example.

\subsection{Type III: Minimal time dispersion of subchannels}
This condition does not properly constitute a conflict as in the previous two cases. Nevertheless, if the subchannels allotted to a vehicle are spread over several subframes, the duration of the signal becomes longer as well as the channel occupancy, implying that less time would remain for other vehicles to broadcast their signals. Therefore, in the event that a vehicle has a greater QoS requirement, it is more suitable to first allocate subchannels in a same subframe before searching for subchannels across several subframes. This issue is particularly very critical in scenarios with high vehicle density. In order to maximize the number of served vehicles, we impose a restriction that forces the subchannels alloted to any vehicle to be confined to a single subframe. Thus, vehicles may be assigned up to $K$ subchannels from any subframe. For any vehicle $v_i$, the following must be satisfied $ (\sum_{u \in \mathcal{R}_{l}} x_{iu}) (\sum_{u' \in \mathcal{R}_{l'}} x_{iu'}) = 0$, for $l \neq l'$ $\forall l, l' = 1, 2, \dots, L$. Note that the expression is non-zero if vehicle $v_i$ transmits in any two subchannels $r_{u}$ and $r_{u'}$ that belong to subframes $l$ and $l'$, respectively. For $N$ vehicles, this can be expressed as
\begin{equation} \label{e3}
	[({\bf I}_{N \times N} \otimes {\bf Q}_{L \times L}^{+}) {\bf x}_s] \circ [({\bf I}_{N \times N} \otimes {\bf Q}_{L \times L}^{-}) {\bf x}_s] = {\bf 0}_{NL \times 1}.
\end{equation}

The matrices ${\bf Q}^{+}$ and ${\bf Q}^{-}$ also entail important information about the allowed and prohibited time dispersion configurations of subchannels. These two matrices will be better understood with an example in Section IV. 

\subsection{Type IV: One-hop inter-cluster subchannel conflicts}
This issue arises when two or more vehicles that are not aware of each other transmit in the same subchannel. Conversely to Type II---where the affected vehicles are only those that transmit concurrently---in this case, signals coming from different vehicles will combine and possibly become undecodable for their counterparts. This is particularly true for those vehicles lying at the intersection of clusters. This kind of phenomenon cannot be counter-measured in decentralized systems such as V2V \textit{mode-4} or IEEE 802.11p \cite{b4}. In V2V \textit{mode-3}, however, it is possible to prevent it by including the right constraints. Thus, for any two vehicles $v_i \in \mathcal{V}^{(j)}$ and $v_{i'} \in \mathcal{V}^{(j')}$ that belong to different intersecting clusters but do not lie at the intersection, i.e. $\{ v_i, v_{i'} \} \not\subset \{ \mathcal{V}^{(j)} \cap \mathcal{V}^{(j')} \}, ~ \forall j \neq j'$, the following must hold: $x_{ik} x_{{i'}k} = 0, ~\forall r_k \in \mathcal{R}$. In general, for $N$ vehicles this can be cast as  
\begin{equation} \label{e4}
	[({\bf H}_{U \times N}^{+} \otimes {\bf I}_{KL \times KL}) {\bf x}] \circ [({\bf H}_{U \times N}^{-} \otimes {\bf I}_{KL \times KL}) {\bf x}] = {\bf 0}_{U \times 1}
\end{equation}
where $U$ is the number of vehicular pairs within one hop, i.e. this precludes the vehicles that lie at the intersections. The matrices ${\bf H}^{+}$ and ${\bf H}^{-}$ bear information about the vehicles that are within one-hop range and could potentially cause conflicts. 

Thus, upon collecting all the conditions (\ref{e1}), (\ref{e2}), (\ref{e3}) and (\ref{e4}), the exact formulation (EF) of the subchannel allocation problem is shown in (\ref{e5}). The problem shown is a $0\slash1$ quadratically constrained linear program with $N + P + U$ constraints. In order to solve (\ref{e5}), a matrix from each of the quadratic constraints should be extracted before using any optimization solver, thus resulting in a very laborious process. In order to facilitate manipulation of these expressions, in the next section we propose a relaxed version that does not affect the solution optimality.
\begin{figure*}[!t]
	\begin{subequations} \label{e5}
		\begin{gather} 
		\begin{align}
		& {\rm max} ~ {\bf c}^T {\bf x} \\
		& {\rm subject~to}~ \nonumber & \\ 
		& {\bf q}_{N \times 1} - {\bm \epsilon}\leq ({\bf I}_{N \times N} \otimes {\bf 1}_{1 \times KL})({\bf c}_{NKL \times 1} \circ {\bf x}_{NKL \times 1}) \leq {\bf q}_{N \times 1} + {\bm \epsilon} \\
		& [({\bf G}_{P \times N}^{+} \otimes {\bf I}_{L \times L}) ({\bf I}_{NL \times NL} \otimes {\bf 1}_{1 \times K}) {\bf x}] \circ [({\bf G}_{P \times N}^{-} \otimes {\bf I}_{L \times L})({\bf I}_{NL \times NL} \otimes {\bf 1}_{1 \times K}) {\bf x}] = {\bf 0}_{PL \times 1} \\
		& [({\bf I}_{N \times N} \otimes {\bf Q}_{L \times L}^{+}) ({\bf I}_{NL \times NL} \otimes {\bf 1}_{1 \times K}) {\bf x}] \circ [({\bf I}_{N \times N} \otimes {\bf Q}_{L \times L}^{-}) ({\bf I}_{NL \times NL} \otimes {\bf 1}_{1 \times K}) {\bf x}] = {\bf 0}_{NL \times 1} \\
		& [({\bf H}_{U \times N}^{+} \otimes {\bf I}_{KL \times KL}) {\bf x}] \circ [({\bf H}_{U \times N}^{-} \otimes {\bf I}_{KL \times KL}) {\bf x}] = {\bf 0}_{U \times 1}.
		\end{align}
		\end{gather}
	\end{subequations}
	\hrulefill
\end{figure*}

\section{Problem Relaxation} 
We propose a surrogate relaxation for the problem in (\ref{e5}) obtained by linearly combining the constraints. It can be noticed that each of the elements in constraints (5c), (5d) and (5e) are mutually independent and therefore their combination does not alter optimality. Accordingly, a relaxed version of each set of constraints can be obtained through a weighted linear combination of their elements. At each side of the constraints (5c), (5d) and (5e), we multiply by ${\bf w}_1 = {\bf 1}_{PL \times 1}^{T}$, ${\bf w}_2 = {\bf 1}_{NL \times 1}^{T}$ and ${\bf w}_3 = {\bf 1}_{U \times 1}^{T}$, respectively. Thereupon, Property 1 and Property 2 are applied in order to obtain the relaxed formulation (RF) shown in (\ref{e6}), where $\widetilde{\bf G}_{N \times N} = [{\bf G}_{P \times N}^{-}]^T {\bf G}_{P \times N}^{+}$, $ \widetilde{\bf Q}_{L \times L} = [{\bf Q}_{L \times L}^{-}]^T {\bf Q}_{L \times L}^{+} $ and $\widetilde{\bf H}_{N \times N} = [{\bf H}_{U \times N}^{-}]^T {\bf H}_{U \times N}^{+} $. Notice that the resultant number of constraints is $N + 3$ compared to $N + P + U$ constraints of the exact formulation in (\ref{e5}). In order to provide a more intuitive understanding of the matrices above, consider the following example for illustration.  
\begin{figure}[!t]
	\begin{mdframed}[] 
		\textbf{Property 1 (Mixed Hadamard product)}\\
		Let ${\bf x} \in \mathbb{R}^{m}$, ${\bf A} \in \mathbb{R}^{m \times n}$, ${\bf B} \in \mathbb{R}^{m \times n}$, and ${\bf y} \in \mathbb{R}^{n}$, then
		\begin{center}
			${\bf x}^T ({\bf A} \circ {\bf B}) {\bf y} = \rm{tr} \{ diag({\bf x})~{\bf A}~diag({\bf y})~{\bf B} \}$
		\end{center}
	\end{mdframed}
	\begin{mdframed}[] 
		\textbf{Property 2 (Product of two Kronecker products)}\\
		Let ${\bf X} \in \mathbb{R}^{m \times n}$, ${\bf Y} \in \mathbb{R}^{r \times s}$, ${\bf W} \in \mathbb{R}^{n \times p}$, and ${\bf Z} \in \mathbb{R}^{s \times t}$, then
		\begin{center}
			${\bf XY} \otimes {\bf WZ} = ({\bf X} \otimes {\bf W})({\bf Y} \otimes {\bf Z}) \in \mathbb{R}^{mr \times pt}$
		\end{center}
	\end{mdframed}
	\vspace{-0.5cm}
\end{figure}
\begin{figure*}[!t]
	\begin{subequations} \label{e6}
		\begin{gather} 
		\begin{align}
		& {\rm max} ~ {\bf c}^T {\bf x} \\
		& {\rm subject~to}~ \nonumber & \\ 
		& {\bf q}_{N \times 1} - {\bm \epsilon}\leq ({\bf I}_{N \times N} \otimes {\bf 1}_{1 \times KL})({\bf c}_{NKL \times 1} \circ {\bf x}_{NKL \times 1}) \leq {\bf q}_{N \times 1} + {\bm \epsilon} \\
		& {\bf x}^T ({\bf I}_{NL \times NL} \otimes {\bf 1}_{K \times 1}) \{ \widetilde{\bf G}_{N \times N} \otimes {\bf I}_{L \times L} \} ({\bf I}_{NL \times NL} \otimes {\bf 1}_{1 \times K}) {\bf x} = 0 \\
		& {\bf x}^T ({\bf I}_{NL \times NL} \otimes {\bf 1}_{K \times 1}) \{ {\bf I}_{N \times N} \otimes \widetilde{\bf Q}_{L \times L} \} ({\bf I}_{NL \times NL} \otimes {\bf 1}_{1 \times K}) {\bf x} = 0 \\
		& {\bf x}^T \{ \widetilde{\bf H}_{N \times N} \otimes {\bf I}_{KL \times KL} \} {\bf x} = 0. 
		\end{align}
		\end{gather}
	\end{subequations}
	\hrulefill
\end{figure*} 

\textbf{Toy example:} Assume that there are $N = 4$ vehicles distributed into $J = 2$ clusters, such that $\mathcal{V}^{(1)} = \{v_1, v_2, v_3\}$ and $\mathcal{V}^{(2)} = \{v_1, v_2, v_4\}$ with $\mathcal{V}^{(1)} \cap \mathcal{V}^{(2)} = \{v_1, v_2\}$. Also, $K = 3$ and $L = 3$. Thus, the matrices $\widetilde{\bf G}$, $\widetilde{\bf Q}$ and $\widetilde{\bf H}$ will contain information about potential conflicts of Type II, Type III and Type IV, respectively, for the described scenario.

\footnotesize
\begin{gather} \nonumber
{\bf G}^{-} = 
\begin{bmatrix}
1 & 0 & 0 & 0 \\
1 & 0 & 0 & 0 \\
1 & 0 & 0 & 0 \\
0 & 1 & 0 & 0 \\
0 & 1 & 0 & 0 
\end{bmatrix}
{\bf G}^{+} = 
\begin{bmatrix}
0 & 1 & 0 & 0 \\
0 & 0 & 1 & 0 \\
0 & 0 & 0 & 1 \\
0 & 0 & 1 & 0 \\
0 & 0 & 0 & 1 
\end{bmatrix}
\widetilde{\bf G} = 
\begin{bmatrix}
0 & 1 & 1 & 1 \\
0 & 0 & 1 & 1 \\
0 & 0 & 0 & 0 \\
0 & 0 & 0 & 0 
\end{bmatrix}
\\ \nonumber
{\bf Q}^{-} = 
\begin{bmatrix}
0 & 0 & 0  \\
1 & 0 & 0 \\
0 & 1 & 1 
\end{bmatrix}
{\bf Q}^{+} = 
\begin{bmatrix}
1 & 0 & 0 \\
1 & 0 & 0 \\
0 & 1 & 0
\end{bmatrix}
\widetilde{\bf Q} = 
\begin{bmatrix}
0 & 0 & 0 \\
1 & 0 & 0 \\
1 & 1 & 0
\end{bmatrix}
\\ \nonumber
{\bf H}^{-} = 
\begin{bmatrix}
0 \\ 
0 \\
1 \\
0 
\end{bmatrix}
{\bf H}^{+} = 
\begin{bmatrix}
0 \\
0 \\
0 \\
1 
\end{bmatrix}
\widetilde{\bf H} = 
\begin{bmatrix}
0 & 0 & 0 & 0 \\
0 & 0 & 0 & 0 \\
0 & 0 & 0 & 0 \\
0 & 0 & 1 & 0
\end{bmatrix}
\end{gather} 
\normalsize

The dimensions of $\widetilde{\bf G}$ are $4 \times 4$ because there are $N = 4$ vehicles with each row and column representing one of them. For instance, notice that a Type II conflict will arise if $v_1$ and $v_2$ transmit concurrently as they both belong to the same cluster, thus $[\widetilde{\bf G}]_{12} = 1$. However, $[\widetilde{\bf G}]_{34} = 0$ because $v_3$ and $v_4$ are in different clusters and the condition would not be violated unless both vehicles transmit in the same subchannel but this will be addressed by matrix $\widetilde{\bf H}$. $\widetilde{\bf Q}$ is also a square matrix of dimensions $3 \times 3$ where each row and column represents a subframe. For example, if any of the vehicles attempts to transmit its signal using subchannels of subframes $l=2$ and $l=3$, a Type III conflict is generated and therefore $[\widetilde{\bf Q}]_{23} = 1$. However, if a  vehicle transmits in subchannels that solely belong to subframe $l=3$, then the condition is satisfied and therefore $[\widetilde{\bf Q}]_{33} = 0$. On the other hand, $\widetilde{\bf H}$ has dimensions $4 \times 4$ where each row and column represents a vehicle. For instance, the condition Type IV will be contravened when two vehicle that are at one-hop distance transmit in the same subchannel. Such case happens for $v_3$ and $v_4$, hence $[\widetilde{\bf H}]_{43} = 1$. ${\bf G}^{+}$ and ${\bf G}^{-}$ have 5 rows because all the possible pairs are $\left\langle v_1, v_2 \right\rangle$, $\left\langle v_1, v_3 \right\rangle$, $\left\langle v_1, v_4 \right\rangle$, $\left\langle v_2, v_3 \right\rangle$ and $\left\langle v_2, v_4 \right\rangle$, thus $P = 5$. For ${\bf H}^{+}$ and ${\bf H}^{-}$ there is only one potential conflicting pair, thus $U = 1$. Each of the matrices ${\bf G}^{+}$, ${\bf G}^{-}$, ${\bf Q}^{+}$, ${\bf Q}^{-}$, ${\bf H}^{+}$, ${\bf H}^{-}$ has its own meaning and are complementary to each other in pairs. Further insights about the meaning of these matrices is left to the reader for interpretation as similar logic used for $\widetilde{\bf G}$, $\widetilde{\bf Q}$ and $\widetilde{\bf H}$ applies to them. Moreover, the matrices $\widetilde{\bf G}$, $\widetilde{\bf Q}$ and $\widetilde{\bf H}$, will either be upper or lower triangular because we only consider the pairs of conflicting vehicles without permutation.
 
\section{Simulations}
In all the simulations, we have considered four clusters such that $\lvert \mathcal{V}^{(1)} \lvert = 16$, $\lvert \mathcal{V}^{(2)} \lvert = 16$, $\lvert \mathcal{V}^{(3)} \lvert = 16$, $\lvert \mathcal{V}^{(4)} \lvert = 8$, and $\lvert \mathcal{V}^{(1)} \cap \mathcal{V}^{(2)} \cap \mathcal{V}^{(3)} \lvert = 8$, $\lvert \mathcal{V}^{(1)} \cap \mathcal{V}^{(4)} \lvert = \emptyset$, $\lvert \mathcal{V}^{(2)} \cap \mathcal{V}^{(4)} \lvert = \emptyset$, $\lvert \mathcal{V}^{(3)} \cap \mathcal{V}^{(4)} \lvert = \emptyset$. Thus, there is a total of $N = 40$ vehicles where 8 of them lie at the intersection of clusters $\mathcal{V}^{(1)}$, $\mathcal{V}^{(2)}$ and $\mathcal{V}^{(3)}$ whereas $\mathcal{V}^{(4)}$ is an isolated distant cluster. In addition, we consider that the QoS requirements of each vehicle can be 12 Mbps, 10 Mbps, 5 Mbps or 3 Mbps. Furthermore, there are 10 vehicles of each kind requiring a different QoS (rate) and these are spread across the four clusters. Now, we examine two different scenarios for distinct values of $K$, $L$, and $\epsilon$.

\underline{\textit{Scenario 1}}: \emph{The number of subframes is $L = 16$ whereas the number of subchannels per subframe is $K = 4$. Also, $\epsilon = 0.8$ Mbps and therefore the range of rates are $[ 11.2 - 12.8 ]$ Mbps, $[ 9.2 - 10.8 ]$ Mbps, $[ 4.2 - 5.8 ]$ Mbps and $[ 2.2 - 3.8 ]$ Mbps.}
\begin{figure}[!t]
	\centering
	\begin{tikzpicture}
	\begin{axis}[
	ybar,
	ymin = 0,
	ymax = 14.9,
	width = 9.2cm,
	height = 4cm,
	bar width = 15pt,
	tick align = inside,
	x label style={align=center, font=\footnotesize,},
	ylabel = {Rate [Mbps]},
	y label style={at={(-0.075,0.5)}, font=\footnotesize,},
	nodes near coords,
	every node near coord/.append style={color = black, font = \fontsize{6}{5}\selectfont},
	nodes near coords align = {vertical},
	symbolic x coords = {Average, Maximum, Minimum, Std. Dev.},
	x tick label style = {text width = 1.6cm, align = center, font = \footnotesize,},
	xtick = data,
	enlarge y limits = {value = 0.3, upper},
	enlarge x limits = 0.18,
	legend columns=1,
	legend style={at={(0.665,0.5)}, anchor=south west, font=\fontsize{6}{5}\selectfont, text width=1.8cm,text height=0.02cm,text depth=.ex, fill = none, }]
	
	\addplot[fill = color0] coordinates {(Average,  12.3) (Maximum, 12.8) (Minimum, 11.2) (Std. Dev., 0.42)}; \addlegendentry{Exact Formulation}
	
	\addplot[fill = color2] coordinates {(Average,  12.3) (Maximum, 12.8) (Minimum, 11.2) (Std. Dev., 0.42)}; \addlegendentry{Relaxed Formulation}
	
	\addplot[fill = color8] coordinates {(Average,  5.15) (Maximum, 14.9) (Minimum, 0.91) (Std. Dev., 3.82)}; \addlegendentry{Random Allocation}
	
	\end{axis}
	\end{tikzpicture}
	\caption{Scenario 1 / Vehicles with QoS = 12 Mbps}
	\label{fig3}
	\vspace{-0.25cm}
\end{figure}
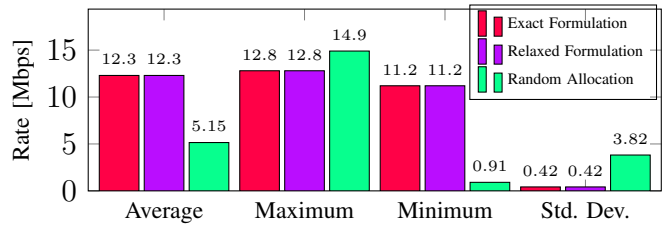
\begin{figure}[!t]
	\centering
	\begin{tikzpicture}
	\begin{axis}[
	ybar,
	ymin = 0,
	ymax = 15.5,
	width = 9.2cm,
	height = 4cm,
	bar width = 15pt,
	tick align = inside,
	x label style={align=center, font=\footnotesize,},
	ylabel = {Rate [Mbps]},
	y label style={at={(-0.075,0.5)}, font=\footnotesize,},
	nodes near coords,
	every node near coord/.append style={color = black, font = \fontsize{6}{5}\selectfont},
	nodes near coords align = {vertical},
	symbolic x coords = {Average, Maximum, Minimum, Std. Dev.},
	x tick label style = {text width = 1.6cm, align = center, font = \footnotesize,},
	xtick = data,
	enlarge y limits = {value = 0.3, upper},
	enlarge x limits = 0.18,
	legend columns=1,
	legend style={at={(0.665,0.5)}, anchor=south west, font=\fontsize{6}{5}\selectfont, text width=1.8cm,text height=0.02cm,text depth=.ex, fill = none, }]

	\addplot[fill = color0] coordinates {(Average,  10.2) (Maximum, 10.7) (Minimum, 9.07) (Std. Dev., 0.39)}; \addlegendentry{Exact Formulation}
	
	\addplot[fill = color2] coordinates {(Average,  10.2) (Maximum, 10.7) (Minimum, 9.07) (Std. Dev., 0.39)}; \addlegendentry{Relaxed Formulation}

	\addplot[fill = color8] coordinates {(Average,  6.36) (Maximum, 15.5) (Minimum, 1.19) (Std. Dev., 4.38)}; \addlegendentry{Random Allocation}
	
	\end{axis}
	\end{tikzpicture}
	\caption{Scenario 1 / Vehicles with QoS = 10 Mbps}
	\label{fig4}
	\vspace{-0.25cm}
\end{figure}
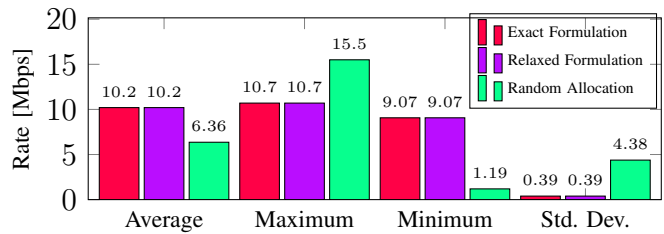
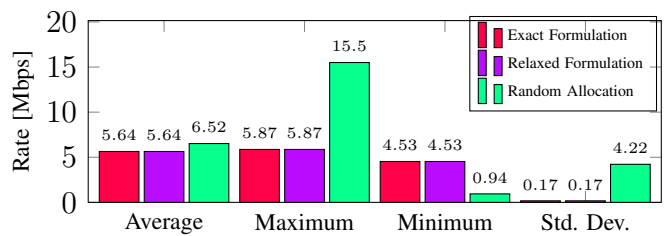
\begin{figure}[!t]
	\centering
	\begin{tikzpicture}
	\begin{axis}[
	ybar,
	ymin = 0,
	ymax = 15.5,
	width = 9.2cm,
	height = 4cm,
	bar width = 15pt,
	tick align = inside,
	x label style={align=center, font=\footnotesize,},
	ylabel = {Rate [Mbps]},
	y label style={at={(-0.075,0.5)}, font=\footnotesize,},
	nodes near coords,
	every node near coord/.append style={color = black, font = \fontsize{6}{5}\selectfont},
	nodes near coords align = {vertical},
	symbolic x coords = {Average, Maximum, Minimum, Std. Dev.},
	x tick label style = {text width = 1.6cm, align = center, font = \footnotesize,},
	xtick = data,
	enlarge y limits = {value = 0.3, upper},
	enlarge x limits = 0.18,
	legend columns=1,
	legend style={at={(0.665,0.5)}, anchor=south west, font=\fontsize{6}{5}\selectfont, text width=1.8cm,text height=0.02cm,text depth=.ex, fill = none, }]
	
	\addplot[fill = color0] coordinates {(Average,  5.64) (Maximum, 5.87) (Minimum, 4.53) (Std. Dev., 0.17)}; \addlegendentry{Exact Formulation}
	
	\addplot[fill = color2] coordinates {(Average,  5.64) (Maximum, 5.87) (Minimum, 4.53) (Std. Dev., 0.17)}; \addlegendentry{Relaxed Formulation}

	\addplot[fill = color8] coordinates {(Average,  6.52) (Maximum, 15.5) (Minimum, 0.94) (Std. Dev., 4.22)}; \addlegendentry{Random Allocation}
	
	\end{axis}
	\end{tikzpicture}
	\caption{Scenario 1 / Vehicles with QoS = 5 Mbps}
	\label{fig5}
	\vspace{-0.2cm}
\end{figure}
\begin{figure}[!t]
	\centering
	\begin{tikzpicture}
	\begin{axis}[
	ybar,
	ymin = 0,
	ymax = 17.94,
	width = 9.2cm,
	height = 4cm,
	bar width = 15pt,
	tick align = inside,
	x label style={align=center, font=\footnotesize,},
	ylabel = {Rate [Mbps]},
	y label style={at={(-0.075,0.5)}, font=\footnotesize,},
	nodes near coords,
	every node near coord/.append style={color = black, font = \fontsize{6}{5}\selectfont},
	nodes near coords align = {vertical},
	symbolic x coords = {Average, Maximum, Minimum, Std. Dev.},
	x tick label style = {text width = 1.6cm, align = center, font = \footnotesize,},
	xtick = data,
	enlarge y limits = {value = 0.3, upper},
	enlarge x limits = 0.18,
	legend columns=1,
	legend style={at={(0.665,0.5)}, anchor=south west, font=\fontsize{6}{5}\selectfont, text width=1.8cm,text height=0.02cm,text depth=.ex, fill = none, }]
	
	\addplot[fill = color0] coordinates {(Average, 3.61) (Maximum, 3.91) (Minimum, 2.58) (Std. Dev., 0.15)}; \addlegendentry{Exact Formulation}
	
	\addplot[fill = color2] coordinates {(Average, 3.61) (Maximum, 3.91) (Minimum, 2.58) (Std. Dev., 0.15)}; \addlegendentry{Relaxed Formulation}
	
	\addplot[fill = color8] coordinates {(Average,  5.95) (Maximum, 17.94) (Minimum, 1.12) (Std. Dev., 4.21)}; \addlegendentry{Random Allocation}
	
	\end{axis}
	\end{tikzpicture}
	\caption{Scenario 1 / Vehicles with QoS = 3 Mbps}
	\label{fig6}
	\vspace{-0.2cm}
\end{figure}
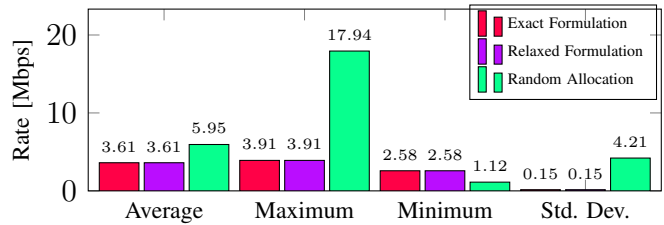

The achieved data rate values for each group of vehicles are shown in Fig. \ref{fig3}, Fig. \ref{fig4}, Fig. \ref{fig5} and Fig. \ref{fig6}. We compare three approaches: the exact formulation (EF) shown in (\ref{e5}), the relaxed formulation (RF) presented in (\ref{e6}) and a random allocation (RA). In the latter case, a randomly generated binary vector ${\bf x}$ does not always satisfy the four types of conditions. Searching for such a solution may sometimes require high computational time. Therefore---for the RA scheme only---Monte Carlo simulations were performed until a feasible solution was found, where only Type II, Type III and Type IV conditions were strictly enforced as each one of them represents a critical conflict. Type I condition is related to QoS demands, which may not be as detrimental as the other types. We examine 4 criteria, \textit{average data rate}, \textit{maximum data rate}, \textit{minimum data rate} and \textit{data rate standard deviation} to assess the performance of the approaches. 

Observe that in all cases, RF attains the same performance as EF, and thus there is no optimality deterioration due to relaxation. Furthermore, in terms of computational time and problem manipulation, RF is simpler and less convoluted than EF. Notice that in both approaches the \textit{maximum data rate} and \textit{minimum data rate} across all the vehicles experience a very narrow deviation from the desired QoS values. This can also be recognized from observing the \textit{data rate standard deviation}, which is very small. If we judged the approaches based on the \textit{average data rate} it may appear that the three of them perform equally fair. However, the performance of RA is inferior as the attained QoS values are either excessively high or insufficiently low compared to the target values. For this scenario, we found that RF is very robust as it does not produce conflicts of any type. However, RA is highly prone to generate conflicts due to its logic-less basis and therefore testing was necessary before accepting a potential solution. As a side note, the value of $\epsilon$ was carefully tuned such that the success rate over 2000 simulations was 100\%. If the QoS ranges are too stringent, a feasible solution may not be found. For instance, the same scenario with $\epsilon = 0.4$ Mbps resulted in a success rate of 88\%.

\underline{\textit{Scenario 2a}}: \emph{Consider that $L = 16$, $K_1 = 3$, $\epsilon_{1} = 1.0$ Mbps and the data rate ranges are $[ 11.0 - 13.0 ]$ Mbps, $[ 9.0 - 11.0 ]$ Mbps, $[ 4.0 - 6.0 ]$ Mbps and $[ 2.0 - 4.0 ]$ Mbps.}  

\underline{\textit{Scenario 2b}}: \emph{Consider that $L = 16$, $K_2 = 7$, $\epsilon_{2} = 0.6$ Mbps and the data rate ranges are $[ 11.4 - 12.6 ]$ Mbps, $[ 9.4 - 10.6 ]$ Mbps, $[ 4.4 - 5.6 ]$ Mbps and $[ 2.4 - 3.6 ]$ Mbps.} 
\begin{figure*}[!t]
	\centering
	\begin{tikzpicture}
	\begin{axis}[
	ybar,
	ymin = 0,
	ymax = 12.9,
	width = 18.5cm,
	height = 3.8cm,
	bar width = 12pt,
	tick align = inside,
	x label style={align=center, font=\footnotesize,},
	ylabel = {Rate [Mbps]},
	y label style={at={(-0.025,0.5)}, font=\footnotesize,},
	nodes near coords,
	every node near coord/.append style={color = black, rotate = 90, anchor = west, font = \fontsize{7}{8}\selectfont},
	nodes near coords align = {vertical},
	symbolic x coords = {QoS$_{\text{12 Mbps}}^{\text{max}}$, QoS$_{\text{12 Mbps}}^{\text{min}}$, QoS$_{\text{12 Mbps}}^{\text{std. dev.}}$, QoS$_{\text{10 Mbps}}^{\text{max}}$, QoS$_{\text{10 Mbps}}^{\text{min}}$, QoS$_{\text{10 Mbps}}^{\text{std. dev.}}$, QoS$_{\text{5 Mbps}}^{\text{max}}$, QoS$_{\text{5 Mbps}}^{\text{min}}$, QoS$_{\text{5 Mbps}}^{\text{std. dev.}}$, QoS$_{\text{3 Mbps}}^{\text{max}}$, QoS$_{\text{3 Mbps}}^{\text{min}}$, QoS$_{\text{3 Mbps}}^{\text{std. dev.}}$},
	x tick label style = {text width = 3cm, align = center, font = \fontsize{8}{9}\selectfont},
	xtick = data,
	enlarge y limits = {value = 0.5, upper},
	enlarge x limits = 0.05,
	legend columns=1,
	legend pos = north east,
	legend style={font=\fontsize{6}{5}\selectfont, text width=1.5cm,text height=0.02cm,text depth=.ex, fill = none, }]
	
	\addplot[fill = color1] coordinates {(QoS$_{\text{12 Mbps}}^{\text{max}}$, 12.4) (QoS$_{\text{12 Mbps}}^{\text{min}}$, 11.6) (QoS$_{\text{12 Mbps}}^{\text{std. dev.}}$, 0.22) (QoS$_{\text{10 Mbps}}^{\text{max}}$, 10.3) (QoS$_{\text{10 Mbps}}^{\text{min}}$, 9.38) (QoS$_{\text{10 Mbps}}^{\text{std. dev.}}$, 0.18) (QoS$_{\text{5 Mbps}}^{\text{max}}$, 5.47) (QoS$_{\text{5 Mbps}}^{\text{min}}$, 5.02) (QoS$_{\text{5 Mbps}}^{\text{std. dev.}}$, 0.05) (QoS$_{\text{3 Mbps}}^{\text{max}}$, 3.51) (QoS$_{\text{3 Mbps}}^{\text{min}}$, 2.84) (QoS$_{\text{3 Mbps}}^{\text{std. dev.}}$, 0.08)}; \addlegendentry{RF / $K_1 = 7$}
	
	\addplot[fill = color6] coordinates {(QoS$_{\text{12 Mbps}}^{\text{max}}$, 12.9) (QoS$_{\text{12 Mbps}}^{\text{min}}$, 11.7) (QoS$_{\text{12 Mbps}}^{\text{std. dev.}}$, 0.26) (QoS$_{\text{10 Mbps}}^{\text{max}}$, 10.9) (QoS$_{\text{10 Mbps}}^{\text{min}}$, 9.51) (QoS$_{\text{10 Mbps}}^{\text{std. dev.}}$, 0.21) (QoS$_{\text{5 Mbps}}^{\text{max}}$, 6.09) (QoS$_{\text{5 Mbps}}^{\text{min}}$, 5.11) (QoS$_{\text{5 Mbps}}^{\text{std. dev.}}$, 0.15) (QoS$_{\text{3 Mbps}}^{\text{max}}$, 4.08) (QoS$_{\text{3 Mbps}}^{\text{min}}$, 3.39) (QoS$_{\text{3 Mbps}}^{\text{std. dev.}}$, 0.11)}; \addlegendentry{RF / $K_2 = 3$}

	\end{axis}
	\end{tikzpicture}
	\caption{Scenario 2 / Attained QoS values for $K_1 = 3$ and $K_2 = 7$}
	\label{fig7}
	\vspace{-0.2cm}
\end{figure*}
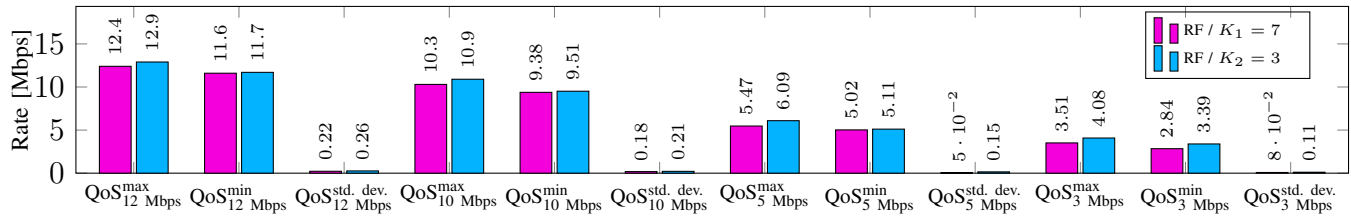
In this scenario we evaluate the performance of the relaxed formulation under both ($i$) very stringent conditions when $K_1 = 3$ and ($ii$) a more relaxed setting when $K_2 = 7$. In the former case, there are $K_1L = 48$ subchannels whereas in the latter case $K_2L = 112$. Also for each of these cases, the value of  $\epsilon $ was chosen such that the success ratio was 100\%. In the latter case the value of $\epsilon$ could be set lower due to the larger amount of subchannels. Hence, due to higher diversity, the requirements could be met more exactly with smaller deviations. In Fig. \ref{fig7}, the maximum and minimum rates for each group of vehicles are shown for the two value of $K$. As can be observed, in both cases the obtained values are within the defined allowable rate bounds. However, in Scenario 2b, the attained values are tighter and closer to the target QoS, which make it more efficient in terms of providing vehicles with the required demands. The values for RA have not been included in Fig. \ref{fig7} due to exaggerated deviations from the desired targets as occurred in the previous scenario.

\underline{\textbf{Note}}: The message rate assumed is 10 Hz implying that each vehicle will have access to the medium every 100 ms. Therefore---in strict sense---the maximum number of subframes is $L_{max} = 100$. Nevertheless, since the number of vehicles in the scenarios is smaller than $L_{max}$ we selected $L$ to be equal to the maximum number of vehicles in all the clusters, i.e. $ L = \max \{ \lvert \mathcal{V}^{(1)} \lvert, \lvert \mathcal{V}^{(2)} \lvert, \dots, \lvert \mathcal{V}^{(J)} \lvert \} $. In this manner, the amount of subframes scales with the maximum number of vehicles assuring the minimum possible value of $L $ that guarantees communication without conflicts. Thus, $L$ subframes are randomly selected among $L_{max}$ available .

\section{Conclusion}
In this work we have presented a mathematical framework for subchannel allocation for vehicular communications V2V \textit{mode-3}. In addition, four types of conditions have been incorporated in order to guarantee a conflict-free allocation that complies with QoS requirements of vehicles. Out of the four conditions three of them are of forceful nature as they depict conflicts in the proper sense. In addition, a relaxed formulation of the exact problem was developed and it was shown through simulations that it does not compromise optimality.

\end{document}